\documentstyle[aps,prl,twocolumn,psfig,url,hyperref,times]{revtex}

\begin{document}

\wideabs{
\title{LSND, SN1987A, and CPT Violation\cite{acknowledgements}}

\author{Hitoshi Murayama$^{1,2}$ and T. Yanagida$^3$}

\address{${}^1$ Theoretical Physics Group,
     Ernest Orlando Lawrence Berkeley National Laboratory\\
     University of California, Berkeley, California 94720}
\address{${}^2$ Department of Physics,
     University of California, Berkeley, California 94720}
\address{${}^3$ Department of Physics and RESCEU, University of Tokyo,
  Hongo, Bunkyo-ku, Tokyo 113-0033, Japan }

\maketitle
\begin{abstract}
  We point out that neutrino events observed at Kamiokande and IMB
  from SN1987A disfavor the neutrino oscillation parameters preferred
  by the LSND experiment.  For $\Delta m^2 > 0$ (the light side), the
  electron neutrinos from the neutronization burst would be lost,
  while the first event at Kamiokande is quite likely to be due to an
  electron neutrino.  For $\Delta m^2 < 0$ (the dark side), the
  average energy of the dominantly $\bar{\nu}_e$ events is already
  lower than the theoretical expectations, which would get aggravated
  by a complete conversion from $\bar{\nu}_\mu$ to $\bar{\nu}_e$.  If
  taken seriously, the LSND data are disfavored independent of the
  existence of a sterile neutrino.  A possible remedy is CPT
  violation, which allows different mass spectra for neutrinos and
  anti-neutrinos and hence can accommodate atmospheric, solar and LSND
  data without a sterile neutrino.  If this is the case, Mini-BooNE
  must run in $\bar{\nu}$ rather than the planned $\nu$ mode to test
  the LSND signal.  We speculate on a possible origin of CPT violation.
\end{abstract}
\pacs{~}
}
\narrowtext

The neutrino masses are strictly zero in the standard model, while
recent strong evidence for oscillations in atmospheric neutrino data
suggests a small but finite mass for neutrinos \cite{atmos}.  There
are also weaker but compelling hints for oscillation in solar neutrino
data \cite{solar}.  Both of them rely on the ``disappearance'' of the
neutrinos compared to theoretical expectations.  On the other hand,
there is a dedicated neutrino oscillation experiment, LSND, which
reported the appearance of $\bar{\nu}_e$ in the $\bar{\nu}_\mu$ flux
from the stopped $\mu^+$ decay \cite{DAR,Mills}.  They have also
reported a hint for appearance in $\nu_\mu \rightarrow \nu_e$ mode but
the significance is low \cite{DIF}.  It was reported that its
significance became even lower in the final analysis \cite{Mills}, and
hence we will ignore this hint throughout this letter.  It is
therefore an important question if all three indications for neutrino
oscillation would fit together.

The observation of neutrinos from SN1987A at Kamiokande and IMB marked
the birth of neutrino astronomy, and confirmed the standard core
collapse model of Type-II supernovae.  Detailed comparisons of data
and theory put constraints on neutrino oscillation parameters (see
\cite{Raffelt} for a review and reference therein).  It is the aim of
this letter to reexamine the constraints from SN1987A neutrino data
with a particular focus on the oscillation parameters preferred by the
LSND experiment.

\begin{figure}[tbp]
  \begin{center}
    \leavevmode
    \psfig{file=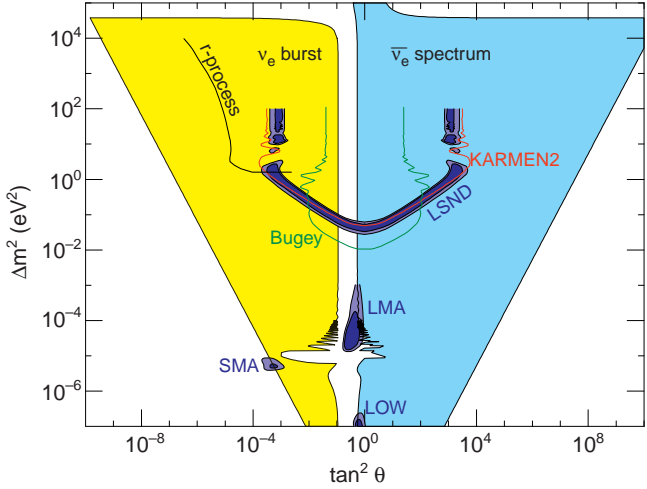,width=\columnwidth}
    \caption{Constraints on the two-flavor oscillation parameter space from
      SN1987A neutrino data.  The shaded MSW triangle on the light
      side ($\tan^2 \theta <1$) is disfavored by the neutronization
      $\nu_e$ burst, while that on the dark side ($\tan^2 \theta > 1$)
      by the energy spectrum of the $\bar{\nu}_e$ induced events.  The
      constraint in order to preserve the nuclear $r$-process excludes
      the region above the curve \protect\cite{Qian}.  LSND preferred
      region is shown at 90 and 99\% CL.  KARMEN2, Bugey constraints
      at 90\% CL are taken from \protect\cite{KARMEN} and exclude the
      regions above the curves.  The currently preferred regions (95\%
      and 99\% CL) from the solar neutrino data \protect\cite{Concha}
      are shown for comparison.}
    \label{fig:constraints}
  \end{center}
\end{figure}

There are basically three types of constraints one can draw from the
SN1987A data.  The first constraint comes from the energy spectrum of
observed events, which are believed to be dominated by $\bar{\nu}_e$
events.  Because of different reaction rates in the proto-neutron star
core, one expects a temperature hierarchy $T_{\nu_e} < T_{\bar{\nu}_e}
< T_{\nu_\mu,\bar{\nu}_\mu,\nu_\tau,\bar{\nu}_\tau}$.  Their average
energies are expected to be 10--12~MeV, 14--17~MeV, and 24--27~MeV,
respectively.  The observed energy spectrum at Kamiokande indicates
that the temperature of $\bar{\nu}_e$ was somewhat on the low side of
the theoretical expectations, with an average energy of 7--14~MeV
\cite{Raffelt}.  If there is an efficient conversion between
$\bar{\nu}_e$ and $\bar{\nu}_{\mu}$ or $\bar{\nu}_{\tau}$, it would
increase the energies of the $\bar{\nu}_e$-induced events, aggravating
the tension between data and theory.  Therefore, oscillation
parameters that would lead to such an efficient conversion are
disfavored \cite{SSB,Jegerlehner}.  The MSW effect via the resonance
occurs when $\bar{\nu}_e$ is heavier than $\bar{\nu}_\mu$ for small
mixing angles as suggested by the LSND data, because the matter effect
due to the charged-current interaction with the electrons would bring
the instantaneous eigenvalue of the Hamiltonian of $\bar{\nu}_e$ state
lower and hence can cause level crossing.  For $\Delta m^2 =
0.1$--1~eV$^2$, as suggested by the LSND data, the conversion is
essentially complete and therefore the SN1987A data disfavor such
parameters.  We use the requirement $P_{osc} < 0.35$ by Smirnov,
Spergel and Bahcall based on this argument \cite{SSB}.  However, the
constraint had not been studied on the dark side $\tan^2 \theta >1$ of
the parameter space \cite{darkside} to the best of our knowledge
\cite{foot1}.  The density profile was taken from \cite{MN} with an
empirical approximation
\begin{equation}
  N_e (r) = \left\{ \begin{array}{ll}
      10^{10} N_A & r < 2.15 \times 10^{-4} r_\odot \\
      0.1 N_A (r/r_\odot)^{-3} \quad & r > 2.15 \times 10^{-4} r_\odot
      \end{array} \right.,
\end{equation}
where $N_A$ is the Avogadro number, and we took $E_\nu \sim 25$~MeV.
The resulting constraint on the oscillation parameter space is shown
in Fig.~\ref{fig:constraints} on the dark side.  The wiggles around
$\Delta m^2 \sim 10^{-5}$~eV$^2$ are due to the Earth matter effect.
Because the Large Magellanic Cloud is seen on the southern sky while
both Kamiokande and IMB detectors reside on the northern hemisphere,
the neutrinos from SN1987A had passed through the Earth, causing
regeneration of $\bar{\nu}_e$ and hence making the constraint weaker.
We approximate the effect using a constant density $N_e \sim 3N_A$ and
$R \sim 10^4$~km.

The second constraint comes from the very first (and possibly the
second) event at Kamiokande.  At the time of core collapse, most of
the protons in the iron core of the progenitor are converted to
neutrons to overcome the Coulomb repulsion, releasing electron
neutrinos.  This is called the neutronization or deleptonization
burst.  Near thermal radiation of all species of neutrinos used in the
first constraint appear only about a hundred milliseconds after the
neutronization burst.  The electron neutrinos dominantly scatter
elastically with electrons in water, and produce highly forward peaked
electrons, while $\bar{\nu}_{e}$ absorption on proton produces a
nearly isotropically distributed positrons.  Indeed the very first
event at Kamiokande points beautifully back at the SN1987A and is
completely consistent with this interpretation.  The expected event
rate of $\nu_e$ is, however, about 0.025 at Kamiokande \cite{SS} and
hence the observation is thanks to an upward statistical fluctuation.
If there is an efficient conversion between $\nu_e$ and $\nu_\mu$ or
$\nu_\tau$, the expected neutral-current event rates due to
$\nu_{\mu,\tau}$ would be about 1/7 of that due to the $\nu_e$ events
which have both neutral-current and charged-current amplitudes, and
hence the observation of one event would be highly unlikely.
Therefore, oscillation parameters that would lead to such an efficient
conversion are disfavored \cite{Arafune,MN}.  We require $P_{osc} <
0.90$, so that the observation of one event is possible within 99\%
CL. The MSW effect via the resonance occurs when $\nu_e$ is lighter
than $\nu_\mu$ for small mixing angles as suggested by the LSND data,
because the matter effect due to the charged-current interaction with
the electrons would bring the instantaneous eigenvalue of the
Hamiltonian of $\nu_e$ state higher and hence can
cause level crossing.  For $\Delta m^2 = 0.1$--1~eV$^2$ suggested by
the LSND data, the conversion is essentially complete and therefore
the SN1987A data disfavor such parameters.  The constraint on the
oscillation parameter space is shown in Fig.~\ref{fig:constraints} on
the left half $\tan^2 \theta < 1$ (the light side).

The third constraint is based on the assumption that the expanding
envelope driven by thermal neutrino wind of exploding supernova is the
site of nuclear $r$-process, synthesizing elements beyond iron.  If
the $\nu_e$ is lighter than $\nu_{\mu}$, there may be an efficient
conversion between the two, and the $\nu_e$ wind would have the
temperature of $\nu_\mu$, {\it i.e.}\/ higher than what it normally
is.  The higher temperature of $\nu_e$ would have a higher cross
section to convert neutrons to protons, where protons would end up
mostly in ${}^4$He and would not participate in producing neutron-rich
nuclei required in the nuclear $r$-process.  This consideration places
a constraint at higher values of $\Delta m^2$ \cite{Qian,more}.  The
constraint derived in \cite{Qian} is shown in
Fig.~\ref{fig:constraints}.

For a comparison, we also show the preferred regions of the parameter
space from the solar neutrino data in Fig.~\ref{fig:constraints},
taken from \cite{Concha}.

The important point is that the oscillation parameters preferred by
the LSND data, which could be on either sides of the parameter space,
are both disfavored by the SN1987A neutrino data, even though the
difficulties in theory of supernova explosion and low statistics in
the data do not allow us to draw a definite conclusion.

So far, our analysis has been within the two-flavor mixing scheme.  
However we know there are three light active neutrinos, and it has 
been argued that we may need even a sterile neutrino state to explain 
LSND, atmoshperic, and solar neutrino data by neutrino oscillations.  
Note that our result does not depend on other oscillation effects, in 
particular whether there exists a sterile neutrino or not.  Let us 
consider the case where all current indications for neutrino 
oscillations, LSND, atmospheric, and solar, are correct and hence 
there is one sterile state.  Because there are three independent 
$\Delta m^2$, there are $3!=6$ ways to order them.  Four of them are 
so-called 3+1 models, where one state is separated by $\Delta m^2_{\rm 
LSND}$ while other three are close to each other separated only by 
$\Delta m^2_{\rm atmos}$ and $\Delta m^2_{\rm solar}$.  These models 
used to be disfavored by the combination of CDHS, CCFR, Bugey, and 
atmospheric neutrino data at SuperKamiokande \cite{bilenkii}, but 
recent reanalysis of the LSND data brought the preferred $\Delta m^2$ 
and $\sin^2 2\theta$ smaller and there opened a small acceptable 
region in the parameter space \cite{Kayser}.  In these models, the 
state $\nu_4$ widely separated from the rest is nearly pure $\nu_s$, 
with small mixing of $\nu_e$ and $\nu_\mu$.  The LSND oscillation is 
explained by the product $\sin^2 2\theta_{\rm LSND} = 4 |U_{e4} U_{\mu 
4}^*|^2$.  Two other models are so-called 2+2 models, where two 
doublets, each responsible for atmospheric and solar neutrino 
oscillations, are separated by $\Delta m^2_{\rm LSND}$.  $\nu_e$ 
($\nu_\mu$) state is almost exclusively in the solar (atmospheric) 
doublet.  Recent SuperKamiokande data disfavors pure $\nu_s$ 
oscillation in both solar and atmospheric data, and therefore we need 
to put $\nu_s$ in both doublets \cite{Smirnov}.  Now we follow how the 
states evolve as the neutrinos exit the proto-neutron star core.  In 
3+1 models, either $\nu_e$ or $\bar{\nu}_e$ crosses the $\nu_s$ state 
first, and the transition between these states is well approximated by 
the two-flavor mixing as studied above.  Therefore either of them is 
nearly completely lost into $\nu_s$.  This strengthens the constraint 
from the neutronization burst, while the constraint from the 
$\bar{\nu}_e$ spectrum is unchanged by the $\nu_s$ component (even 
though the overall normalization gets further suppressed).  In 2+2 
models, either $\nu_e$ or $\bar{\nu}_e$ crosses the atmoshperic 
doublet first and are nearly completely converted.  Therefore the 
constraints discussed in the two-flavor case are unaffected.

A fair question to ask is how robust these constraints are.  As for
the temperature difference used in the first constraint, the issue had
been raised if an additional process, such as $\nu N N \rightarrow \nu
N N$, may reduce the temperature differences between $\bar{\nu}_{e}$
and $\nu_{\mu}$, $\bar{\nu}_{\mu}$, $\nu_{\tau}$, $\bar{\nu}_{\tau}$
\cite{Hannestad}, but no concrete estimates of the temperature had
been given.  This issue can be settled only by more detailed numerical
simulations and/or a future observation of supernova neutrino bursts.
For instance, SuperKamiokande, SNO, Borexino, and KamLAND can detect
$\bar{\nu}_{e}$ via the charged-current reaction, while $\nu_{e}$ can
also be detected at SNO via the charged-current, and all neutrino
species at SNO, Borexino, and KamLAND via the neutral-current reaction
(see \cite{Scholberg} for a recent review on the experimental
aspects).  Then we can test if there is a significant temperature
difference among different event categories.  The interpretation that
the first Kamiokande event, produced at the angle $18^{\circ}\pm
18^{\circ}$ in the forward direction, is due to the elastic $\nu_{e}
e$ scattering of $\nu_{e}$ from the neutronization burst is also
subject to a criticism.  The expected event rate is low, and we are
relying on a single event to place the constraint.  The probability
that this event is due to an isotropically distributed $\bar{\nu}_{e}$
event is about 3\% \cite{SS}.  We find the fact that this event was
the {\it first}\/ quite suggestive of being a $\nu_{e}$ event.  Again
a future detection of supernova neutrinos would settle this issue.
Because of these possible criticisms, we cannot make a definite claim
that the LSND data is incompatible with the SN1987A neutrino events.
We can only say that the LSND preferred region is disfavored by
SN1987A data based on the assumptions made above.

For the rest of the letter, we take the above constraints seriously, 
and we discuss how we may accommodate the LSND data despite the
constraints.  The only way to evade the SN1987A constraints is to 
assume $\nu_e$ is heavier than $\nu_{\mu,\tau}$ while $\bar{\nu}_e$ 
lighter than $\bar{\nu}_{\mu,\tau}$.  Such a mass spectrum obviously 
violates CPT, but we do not see any other alternatives as long as we 
take the SN1987A constraints seriously.  Once CPT is violated, in 
principle one may also consider the violation of Lorentz invariance, 
which will be discussed elsewhere.  For a phenomenological exercise, 
we consider different mass spectra for neutrinos and anti-neutrinos 
and keep Lorentz invariance.  The question is if we can accommodate 
atmospheric, solar, and LSND data within the SN1987A constraints.

\begin{figure}[htbp]
  \begin{center}
    \leavevmode
    \psfig{file=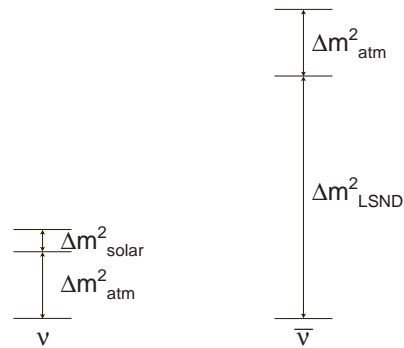,width=0.6\columnwidth}
    \caption{Possible mass spectra of neutrinos and anti-neutrinos
    consistent with solar, atmospheric, LSND data and the SN1987A
    constraints. }
    \label{fig:spectra}
  \end{center}
\end{figure}

The key to this question is that the solar neutrino data probe only
$\nu_e$, but not $\bar{\nu}_e$, while the LSND data only
$\bar{\nu}_e$, but not $\nu_e$ \cite{foot2}.  On the neutrino
spectrum, unless $U_{e3}$ element is extremely small, $\nu_e$ cannot
be below the atmospheric neutrino mass gap because it would cause a
loss in the neutronization burst.  Therefore, the solar mass gap must
be above the atmospheric mass gap.  On the anti-neutrino spectrum, we
need $\bar{\nu}_e$ the lightest, and $\bar{\nu}_\mu$, $\bar{\nu}_\tau$
about 1~eV$^2$ above $\bar{\nu}_e$.  The splitting between two mass
eigenstates which are dominantly $\bar{\nu}_\mu$, $\bar{\nu}_\tau$
must be relevant to atmospheric neutrino oscillation.  The mass
spectra are depicted in Fig.~\ref{fig:spectra}.  It is interesting to
note that the combination of the cosmic microwave background
anisotropy from Planck satellite and Lyman-$\alpha$ power spectrum
will be able to exclude the neutrino mass down to 0.29~eV at 2$\sigma$
level \cite{Croft}.

Different mass spectra between neutrinos and anti-neutrinos will 
affect future neutrino oscillation experiments.  The most important
consequence is for the Mini-BooNE experiment, which is supposed to put
a final word on the LSND signal.  They will run primarily in the
$\nu_\mu$ mode, unfortunately, which would not exhibit the LSND
oscillation.  They do have a capability to run in the
$\bar{\nu}_\mu$ mode, however, and this mode must be used to test the
LSND evidence. 

Brief comments on the possible origin of CPT violation are in order.
First of all, it has been often argued that the small mass of
neutrinos could well be originated from Planck-scale physics.  Even
though Yukawa couplings suppressed by the nominal Planck $h \sim
v/M_{Pl}$ would be too small compared to the required mass spectra
above, the ``Planck scale'' can well be much lower, even down to the
TeV scale as has been discussed intensively lately \cite{ADD}.
Therefore, it is quite possible that the small neutrino masses probe
quantum gravity physics.  It is then also conceivable that the
possible violation of CPT from quantum gravitational physics appears
most evidently in the neutrino mass spectra but not elsewhere.  For
instance, non-commutative geometry violates Lorentz invariance at
short distances, producing possible seeds for CPT violation
\cite{Itzhaki} (see, however, \cite{Sheikh-Jabbari}).  It is easy to
write down Hamiltonian with different massees for neutrinos and
anti-neutrinos in momentum space, but it is non-local in the
coordinate space.  See Refs.~\cite{CPT} for recent discussions on
other stringy or quantum-gravitational origin of CPT violation.  Even
though this discussion is highly speculative, we hope that our work
provokes more intensive discussions on the possible origin of CPT
violation.

In summary, we discussed the SN1987A constraints on the neutrino
oscillation parameters, and found that the parameters preferred by the
LSND data are disfavored by the SN1987A data on both sides of the
parameter space.  If we take these constraints seriously, the only way
to make the LSND data compatible is to allow different mass spectra
for neutrinos and anti-neutrinos, and hence CPT violation.

\acknowledgements{HM thanks Nima Arkani-Hamed, Lawrence Hall, Aaron
  Pierce, and Dave Smith for useful comments.}


\begin{thebibliography}{99}
\bibitem{foot1} The convention adopted here is that the mass
  eigenstate closer to the electron neutrino state is lighter in the
  light side $\tan^2 \theta < 1$ while heavier in the dark side
  $\tan^2 \theta > 1$.
\bibitem{foot2} As mentioned earlier, the
  LSND hint for $\nu_\mu \rightarrow \nu_e$ in decay-in-flight data is
  statistically not significant and we do not consider it in this
  letter.
\end{thebibliography}
\end{document}